# Fractal-Based Heuristic Pixel Segmentation for Lossless Compression of High-Bit-Depth DICOM Medical Images

Taaha Khan[1]

[1] Sunset High School, Portland OR 97229, USA
`taahajkhan@gmail.com`

**Abstract.** Medical image compression is a widely studied field of data processing due to its prevalence in modern digital databases. This application requires a high color depth of 12 bits per pixel component for accurate analysis by physicians, primarily in the DICOM format. Standard raster-based compression of images via filtering is well-known; however, it remains suboptimal in the medical domain due to non-specialized implementations. This study proposes a lossless medical image compression algorithm, CompaCT, that aims to target spatial features and patterns of pixel concentration for dynamically enhanced data processing. The algorithm employs fractal pixel traversal coupled with a novel approach of segmentation and meshing between pixel blocks for preprocessing. Furthermore, delta and entropy coding are applied to this concept for a complete compression pipeline. The proposal demonstrates that the data compression achieved via CompaCT's fractal segmentation preprocessing yields enhanced image compression results while remaining lossless in its reconstruction accuracy. CompaCT is evaluated in its compression ratios on 3954 high-bit-depth CT scans against the efficiency of industry-standard compression techniques (i.e., JPEG2000, RLE, ZIP, PNG). Its reconstruction performance is assessed with error metrics to verify lossless image recovery after decompression. The results demonstrate that CompaCT can compress and losslessly reconstruct medical images, being 37% more space-efficient than industry-standard compression systems.

**Keywords:** Fractal, Image Compression, Medical Image, Segmentation.

## 1 Introduction

### 1.1 Overview

In recent years, the medical domain has witnessed a substantial transition towards digital storage of images in databases worldwide [1]-[2]. While this shift has undoubtedly facilitated data accessibility and exchange, it has also brought to the forefront an inherent challenge: the consumption of significant storage space by large quantities of high-bit-depth medical scans [1]-[3].

    Medical image modalities, such as Computed Tomography (CT), Magnetic Resonance Imaging (MRI), and X-ray, play a pivotal role in medical analysis, supporting

crucial diagnostic and treatment decisions [3]-[6]. Due to their importance, these medical modalities demand more stringent data format requirements. Typically, medical images are stored using 12 bits per color component per pixel, translating to 2 bytes of data per pixel [5], [7]. This contrasts with general images that typically employ 8 bits per color component per pixel [8]. High bit-depth storage for medical images results in larger file sizes, escalating storage costs when preserving these data archives indefinitely [2], [4]-[5].

To address the pressing storage challenge, image compression algorithms have emerged as a viable solution to reduce the required storage space [2], [4]-[5], [8]. Image compression represents a widely studied and extensively utilized field within digital image processing, encompassing a range of techniques and methodologies aimed at minimizing the data size while preserving essential information [2].

Two fundamental forms of data compression exist: lossy and lossless [8]-[9]. Lossy compression reduces file sizes by discarding certain non-essential information from the image and retaining only the key aspects necessary for perceptual data reconstruction [8]-[9]. However, in medical information, lossy algorithms are avoided due to their potential to compromise the integrity of stored data, which is critical for precise and reliable analysis [2]-[5], [8], [10]-[11]. Instead, the medical image domain predominantly relies on lossless compression, which reduces file sizes while ensuring that the compressed format can perfectly reconstruct the original information [8]. This approach retains all the data in the initial image, effectively striking a balance between efficient storage and data fidelity [4].

## 1.2 Digital Imaging and Communications in Medicine (DICOM)

The Digital Imaging and Communications in Medicine (DICOM) format is the primary industry standard for saving and transmitting digital medical images worldwide [12]. DICOM ensures comprehensive and standardized information exchange across medical imaging devices and healthcare institutions by encapsulating image data and accompanying metadata [12]. Medical datasets are commonly stored in the DICOM format, identifiable by the "dcm" extension, facilitating the seamless integration of images into medical information systems, such as in Fig. 1.

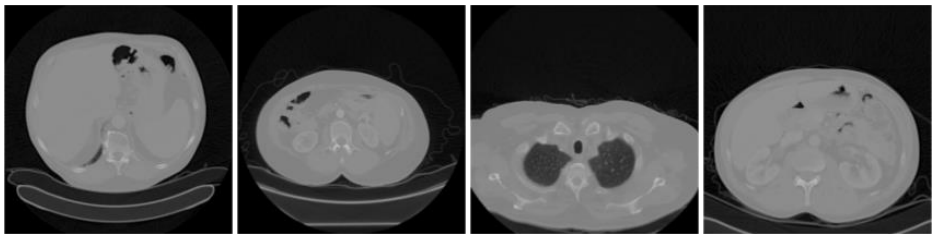

**Fig. 1.** Examples of high-bit-depth CT scans stored in raw DICOM format.

The DICOM standard includes built-in lossless compression formats to facilitate efficient data storage and transmission [12]. These compression algorithms include JPEG2000-Lossless (JP2), Run Length Encoding (RLE), and ZIP [12], with JP2 being the primary industry standard for medical images [3], [6]. However, it is essential to recognize that these built-in compression options are predominantly tailored to handle general-purpose images [7], potentially limiting performance when dealing with dedicated medical images [10].

This limitation underscores the need for a dedicated image compression system to optimize DICOM medical images' efficiency [10]. By harnessing the unique characteristics of medical image modalities, such a specialized system can unlock untapped compression potential and yield significant advancements in storage space optimization, thereby reducing storage costs and facilitating seamless data sharing among healthcare providers.

The proposed medical image compression algorithm, CompaCT, represents an approach focused on pixel restructuring to optimize compression efficiency with a highly specialized segmentation system. Unlike conventional pixel-based compression systems such as Portable Network Graphics (PNG) [13] and ZIP DEFLATE [14], which employ raster-based scans with primarily horizontal patterns during encoding, CompaCT seeks to explore innovative pixel orderings that facilitate enhanced compression performance.

## 1.3 Previous Work

Previous research specifically in pixel restructuring for image compression has investigated various ordering techniques. While some techniques may not result in perfect lossless compression, it is important to develop and understanding of the previous techniques used for pixel restructuring before compression. The primary proposal of the current paper is to develop and enhanced pixel reorganization technique that allows for optimized compression when utilized in tandem with other compression procedures.

Some studies [15]-[16] exploring fractal patterns aim to organize pixels into both vertical and horizontal sections, as well as spatially, to achieve improved compression results. Other research efforts have also experimented with dynamically reconstructed data [17], aiming to enhance compression rates; however, these approaches can overlook some of the specific challenges posed by the high-bit-depth nature of medical information.

Work by Shen and Rangayyan [4] started with segmentation-based medical image coding concepts for high-bit-depth data. Regional flood fill procedures were used to adapt the scanning order into an enhanced image compression codec. This research resulted in about 28% improved performance compared to JPEG on the database used. Similarly, Min and Sadleir [11] proposed a region of interest (ROI) based segmentation approach, but this entailed entirely deleting the image's background. This, while improving compression, can spark concerns about the accuracy of the ROI detection with possibly essential information being discarded [3]. Chen and Wang [9] did similarly with content-based foreground-background segmentation for near-lossless compression, achieving increased compression, but still falling under the same ROI accuracy concerns.

The work of block-based compression by Kumar et al. [18] proposed 4x4 block pixel groupings to aid in spatial pattern recognition and utilization. This research also applied bitwise optimization of encoding integers based on predetermined bounds. This improved compression of several standards, including JP2, but was limited to 8-bit grayscale mammography scans.

This present study proposes a novel pixel restructuring technique explicitly tailored to encode monochrome images stored at 12 bits per pixel, the standard format for many medical image modalities. CompaCT takes advantage of optimized pixel orderings, delta coding, and entropy coding techniques to achieve compression gains while ensuring the preservation of critical diagnostic information.

### 1.4 Overview

The following is a summary of the overall CompaCT pipeline. Initially, the 12-bit-depth image is reordered in a Hilbert fractal scan. This allows for enhanced regional correlations between pixels spatially, as opposed to applying a standard raster scan through the pixel rows. Following the fractal transform, the image is dynamically restructured in a novel proposal. The pixel stream is segmented into 16-pixel blocks, and the blocks are flagged if they contain many large differences. A large difference is defined if it takes more than a single byte to encode the delta between a pixel and it's predecessor in the fractal scan. If the block contains many large differences, future unencoded blocks in the array are searched. Each block is meshed with a potential partner, such that the pixels are interlaces between the two individual blocks and encoded together. If the paired partners result in fewer large differences, then they are grouped together in the complete pixel restructuring. With the restructured data, delta coding is applied to the transformed pixels such that differences that can be encoded within a single byte are encoded, and large differences are encoded using 2 bytes. Following the short byte removal and large delta encoding, the standard DEFLATE backreferencing and entropy coding procedure is utilized for a complete compression pipeline. Each stage in the CompaCT algorithm is losslessly invertible, resulting in perfect image reconstruction after decompression.

The primary novelty is the dynamic pixel segmentation procedure that is utilized following the fractal transform and before the delta and entropy coding systems. This highlights the core contribution of this proposal to the field of lossless medical image compression. The paper is focused on lossless medical image compression as it is a highly important field of data processing, with very specific data requirements in the 12-bit monochrome configuration. With the general format of black backgrounds and a central mass of data in the scan, the segmentation approach focuses on restructuring the edges of the mass in the scan to a more compressible format.

## 2 Methods

### 2.1 Fractal Transform

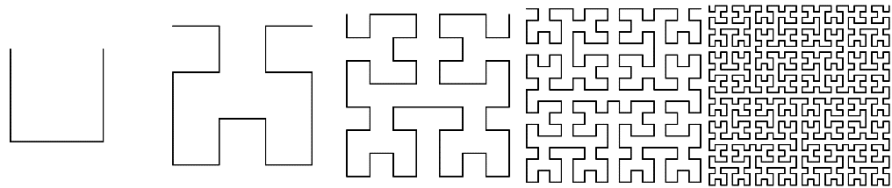

**Fig. 2.** First five iterations of a recursive Hilbert curve generation through a square array. This ordering is applied to iterate through pixels in an image. The algorithm in this proposal generates a dynamic curve suitable for rectangular arrays as well.

The outcome of using a raster scan means that horizontal pixel patterns can be utilized for compression [16]. Some encoders work around this by applying several possible filters, such as PNG [13], for example, applying filters of left pixels, above pixels, and combinations. This allows for two-dimensional patterns to be utilized in the compression [15].

The CompaCT study does not utilize different pixels as candidates for predictive coding. Later steps in this pipeline only relate pixels to the immediate previous pixel. Applying a raster scan with these limited functionalities will result in only horizontal patterns being utilized, creating suboptimal performance. To enhance the different multidimensional patterns in medical image data, a Hilbert-curve [Fig. 2] by Cerveny, 2021 fractal pixel ordering is employed [19]. The specific implementation of the Generalized Hilbert curve is not the primary novelty proposal in this paper, but it is a core step utilized in the transformation procedure to enhance compression performance.

Hilbert fractal pixel ordering is based on incorporating all directions of relationships between pixels using a recursive and coiling fractal structure of the iteration [15]-[16]. The utilization of linearized image coding can reduce correlation in the pixel data compared to standard raster scans [20]. This acknowledges horizontal, vertical, and regional interpixel relationships to be encoded in the data, allowing for possible dynamic enhancements to this field of image preprocessing [15]-[16].

### 2.2 QOI Byte Collapse Introduction

The next encoding step is based on an open-source encoding format called "QOI: The Quite OK Image Format" [21]. The core concepts of QOI were to collapse RGB pixels into smaller byte representations quickly without any entropy coding. In the intended cases, it would collapse many 8-bit color components into one or two bytes by applying run length coding, color caching, and short delta coding. The CompaCT proposal focuses mainly on the short delta compression method and applies it to the domain of high-bit-depth data for monochrome channels in the medical image domain.

6The goal of this QOI-based delta encoding system is to collapse pixel deltas that would have previously taken multiple bytes to encode down into a single byte. This is done by maximizing the number of smaller deltas between pixels and minimizing encoding full multi-byte deltas. The future meshing and restructuring phase aids significantly with this goal of enhancing compression.

When the delta between two 12-bit pixels is encoded, the worst case is that 12-bits are needed to represent that signed integer difference. With the convenience of keeping all pixels byte-aligned, 4 bits must be padded to make the 12 bits of pixels into two complete bytes. On the other hand, encoding a difference that falls between the values -64 and 65 inclusive can be encoded using only a 7-bit signed integer, allowing for a single flag bit. So, if the difference between two adjacent pixels is within this "small difference" range, then a single byte is sufficient to encode the data instead of two bytes for "large differences." Minimizing the total amount of "large differences" when encoding the differences in pixels will, in turn, decrease the number of bytes needed with this delta encoding system.

### 2.3    Count of Large Differences (CLD) Heuristic

The goal of dynamically reordering the pixels in the image is to enhance the steps of delta encoding and entropy coding. The order of pixels directly affects the repetitions that can be utilized in the delta codes. Reorganizing into less entropic data allows for enhanced compression in later stages of this proposed pipeline. The concept is to minimize the amount of "large differences" as defined by the heuristic:

$$L(\Delta) = \begin{cases} 1, & \Delta \leq -64 \\ 0, & -64 < \Delta < 65 \\ 1, & \Delta \geq 65 \end{cases} \quad (1)$$

Equation (1) defines a Large Difference (L) with *delta* being the difference between consecutive pixel intensities. Confining many 12-bit values into single bytes aids in collapsing the number of bytes needed to represent the data without any traditional entropy coding yet and increases the chances of repeating codes given the smaller possibilities of integers. Getting the number of large differences across a series of pixels is calculated:

$$CLD(p) = 1 + \sum_{i=1}^{n} L(p_i - p_{i-1}) \quad (2)$$

Equation (2) defines the Count of Large Differences (CLD) heuristic measuring the number of large differences found over a series of consecutively encoded pixels, *p* being a pixel array, *n* being the number of pixels, and *L* being the large difference as defined in equation (1). The equation adds 1 to the summation result to include the presence of the initial pixel $p_0$.

## 2.4 Pixel Segmentation

The proposed CompaCT system identifies 16-pixel blocks in the Hilbert fractal as single units of pixels, similar to the base proposed by Kumar et al. [18]. The segmentation system traverses through these 16-pixel blocks in order of the iteration produced in the fractal scan.

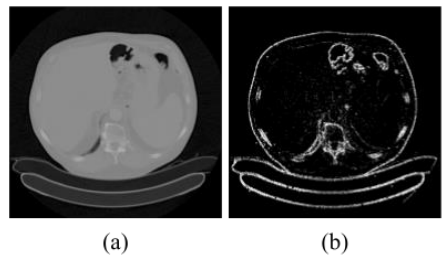

(a)           (b)

**Fig. 3.** The original image (a) is shown side-by-side with identified "difficult" large differences (1) in image (b). White in (b) notates that the difference when encoding that pixel in the fractal order and the previous pixel results in a delta that falls out of the easy delta range of (-64, 65). The large deltas are primarily focused around the edges of the central mass that is present in medical data.

We iterate through the 16-pixel blocks and count the number of large differences (1) present in each delta between the pixels within the block. The pixels within each block are iterated in the full fractal ordering that was generated at the start of the pipeline, meaning that even within the blocks the pixels are not in raster order. In forms of square images, the 16-pixel blocks will be fall into 4x4 shaped tiles, but not always in rectangular cases. An arbitrary cutoff threshold is set to flag the block as "difficult" if more than half of the differences between the pixels fall outside the range of a 7-bit signed integer, as depicted in Fig. 3b. This flag highlights which blocks should be considered when searching for a pair to reorder with, speeding up searching by masking away already simple to compress sections. The CLD (2) of each block can be efficiently computed by initially setting up a prefix sum array to count the number of large deltas in a certain range. The prefix sum array can be generated in linear time, then each query of the CLD on any range of consecutive pixels can be retrieved in constant time.

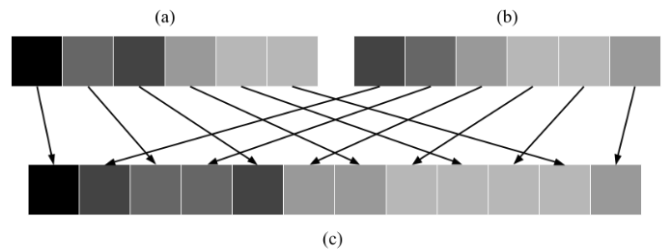

**Fig. 4.** Miniaturized example of interlacing similar blocks (a) and (b) into conjoined block (c).

Once all "difficult" blocks are identified, the system will find pairs of blocks to make efficient meshing partners. A mesh is defined in Fig. 4 as interlacing the pixels in each block by alternating the order in which they are encoded. A mesh of two 16-pixel units will result in an interlaced section of 32 pixels, in which every other pixel belongs to two distinct units.

It is only efficient to interlace blocks together if the resulting mesh creates more common and smaller deltas between pixels, which can more efficiently be minimized by encoding in backreferencing and entropy systems later.

The searching phase iterates through the blocks in the order of the fractal traversal. Once settling on a block that was previously flagged as "difficult," it executes a linear search through future blocks to find a match that will decrease the CLD within the potentially meshed block. The search starts at the next block and continues through the immediate next 64 unencoded blocks. It simulates meshing each block with the original flagged block together and calculates the resulting CLD when the block pixels are interlaced. Suppose the resulting CLD proves to be less than the sum of the original CLD of encoding each block separately. In that case, the encoder meshes the blocks into a single unit.

The next 32 pixels in the pixel stream will be out of order but will enhance the efficiency for later phases of the CompaCT pipeline. In order to identify the segmented and reorganized blocks, a single byte is inserted into the byte stream to identify the location of the future block that was taken to mesh with the current block. With the introduction of this flagging byte to the stream, the meshed block must be able to save at least a single byte of storage by merging multi-byte deltas into single bytes to be relatively efficient. If a single byte or more cannot be saved by meshing two prospective blocks, then the pair is discarded from consideration. This is determined by retrieving the individual expected bytes that would be needed for encoding the current and default next block and comparing it to the prospective meshed block. If the size of the default block bytes is only a single byte shorter or the same length, then meshing the pair is futile and is discarded.

Once a block has been selected to mesh with a previous block, it is removed from later search iterations to avoid redundantly coding the same pixels in multiple sections of the byte stream. The result of the reorganized data results in a stream of pixels in fractal order, along with intermittent meshed flags for the reorganization of blocks. This phase results in a slight data expansion of a few hundred bytes before the actual compression and collapse of bits are done.

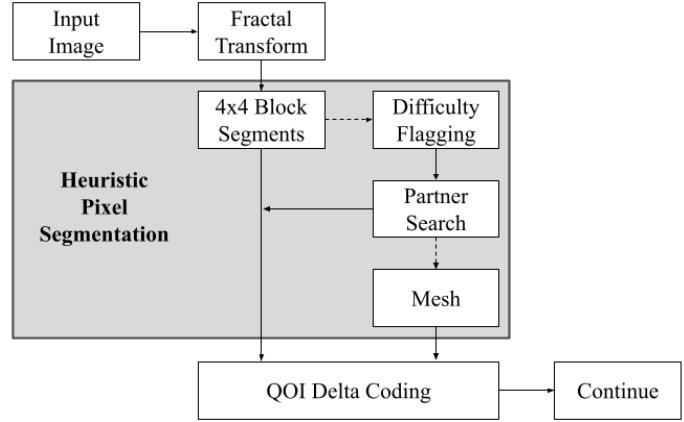

**Fig. 5.** Preprocessing CompaCT image transformation pipeline before continuing encoding.

The pixel restructuring process employed by CompaCT, as defined in Fig. 5, aims to identify and exploit inherent patterns and redundancies within medical images, allowing for more effective data representation. By strategically rearranging pixels, the algorithm can identify local variations in intensity and efficiently represent them using delta coding, which stores the difference between neighboring pixel values rather than their absolute values. This approach capitalizes on how medical images can exhibit gradual intensity transitions in localized regions [9].

### 2.5 QOI-Based Delta Coding

**Table 1.** QOI-based delta coding byte packet metadata.

| Type | Tag Bits | Payload Bits | Tag | Payload Range | Payload Content |
|---|---|---|---|---|---|
| Short Delta | 1 | 7 | 0 | [-64, 65] | Difference from the previous pixel. |
| Meshed Flag | 2 | 6 | 10 | [0, 64] | Block jumps forward of meshed block. |
| Full Delta | 4 | 12 | 1110 | [-2047, 2048] | Difference from the previous pixel. |

With the reorganized pixels into more efficient smaller differences, the QOI system will collapse small differences from two bytes into single bytes. The data packets are labeled with tag bits, as defined in Table 1, for labeling in decoder reconstruction. The tags are customized for this application and no longer resemble the original QOI tags.

### 2.6 DEFLATE Entropy Coding

Furthermore, CompaCT leverages entropy coding techniques to compress the delta-coded pixel data more efficiently. Entropy coding exploits the varying probabilities of pixel values occurring within the image, assigning shorter codes to frequently occurring pixel values and longer codes to less frequent ones [8]. This adaptive coding mechanism enables CompaCT to achieve higher compression ratios by allocating shorter codes to the most common pixel intensities found in medical images.

After processing the reorganized pixels into a QOI byte stream, the information can be further compressed losslessly using standard encoding algorithms. The DEFLATE [14] system is a common final step in many data processing pipelines to result in maximally compressed data.

The DEFLATE procedure applies Lempel-Ziv's (LZ77) backreferencing to simplify commonly repeated phrases in the data [14]. The LZ77 algorithm uses references to previous strings of repeated bytes to minimize repetitions in the data. The reorganized data and minimized deltas encoded in the stream using the fractal-segmentation system aid in creating highly redundant codes that can be shrunk down with LZ77.

The next step in DEFLATE is the Huffman entropy coding system [14], which minimizes the bit representations of the data dynamically based on usage probabilities [8]. This will result in a highly entropic and minimized version of the final data that can be losslessly inverted back into the QOI-based byte stream.

Once the actual pixel data is compressed, the data is written to a resulting output file along with some header bytes containing specific information about the data, including magic characters "pact", image dimensions, number of channels, number of bytes per channel, boolean flag for fractal transform, boolean flag for segmentation transform, and a boolean flag for DEFLATE compression. A decoder can parse this initial information and read the file based on the desired configurations of the compressed input.

### 2.7 Overall pipeline

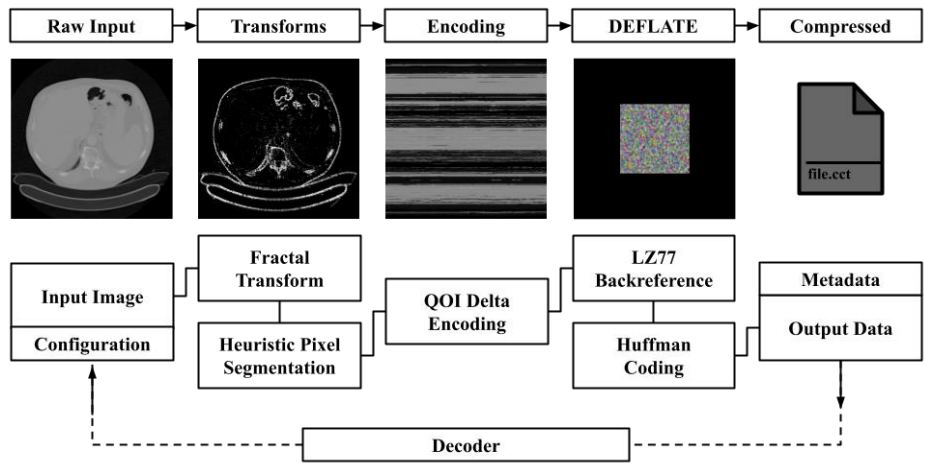

**Fig. 6.** Encoding process of the overall encoding and decoding compression pipeline. Each step is reversible in the decoder algorithm for image reconstruction.

**Algorithm 1: CompaCT Encoding Pipeline Pseudocode**

```
Function COMPACT_ENCODE(image, config):
    output = [] // Initialize an empty sequence called output to store the encoded data.
    WriteHeader(output, config) // Write the header information based on the configuration.

    If config.fractalEnabled:
        fractalOrdering = GenerateFractalOrdering(image) // Generate a fractal ordering
        image = RearrangeImage(image, fractalOrdering) // Rearrange the with fractal

    If config.segmentationEnabled:
        pixelData = FlattenImage(image) // Flatten the image into a 1-dimensional array
        blocks = DivideIntoBlocks(pixelData, config.blockSize) // Divide pixels into blocks

        For each blockIndex from 0 to blocks.length - 1:
            currentBlock = blocks[blockIndex]
            nextBlock = blocks[blockIndex + 1]
            If CalculateCLD(currentBlock) > config.blockSize / 2:
                For each subsequentBlockIndex from blockIndex + 1 to blockIndex + 64:
                    meshedBlocks = MeshBlocks(currentBlock, blocks[subsequentBlockIndex])
                    If CLD(currentBlock + nextBlock) > CLD(meshedBlocks) - 1:
                        UpdatePixelOrder(currentBlock, nextBlock)
            Else:
                KeepDefaultOrdering(currentBlock)

    pixelOrder = GetPixelOrder(blocks) // Get the final order of pixels.

    For each pixel in pixelOrder:
        If IsJumped(pixel):
            InsertJumpLabel(output)
        delta = CalculateDelta(pixel, previousPixel)
        If -64 < delta < 65:
            InsertShortDelta(output, delta)
        Else:
            InsertFullDelta(output, delta)

    CompressedOutput = DEFLATE(output) // Compress using the DEFLATE algorithm.

    Return CompressedOutput
```

The CompaCT pipeline, as outlined in Fig. 6 and Algorithm 1, offers a comprehensive solution for optimizing compression ratios in medical image processing by combining pixel restructuring, delta coding, and entropy coding. This specialized approach considers the unique characteristics of medical images, which demand both high bit-depth representation and preservation of detailed information critical for accurate medical analysis.

## 3 Results

### 3.1 Reconstruction Validation

This study verifies the lossless nature of the CompaCT pipeline by confirming equality when comparing the original and reconstructed images. Lossless compression is critical to medical data reconstruction, and there is no option in the library for any lossy functionality.

### 3.2 Compression Ratio Metric

The compression ratio is a common metric for quantifying an algorithm's compression power [8].

$$CompressionRatio(I) = \frac{UncompressedSize(I)}{CompressedSize(I)} \quad (3)$$

Equation (3) defines the compression ratio of image $I$ to evaluate how many times smaller the compressed image is compared to the size of the original data. The uncompressed size of the file is divided by the resulting compressed size to quantify how much the data was compressed. For example, a compression ratio of 2 corresponds to a 50% smaller file than the original. It is standard practice in the field of lossless image compression to represent algorithm power using the compression ratio metric, even if there are empty bits of 12-bit images of 2-byte representations. The overall compression ratio is reported for analysis and comparison to other papers and reports in the field.

### 3.3 Evaluation Dataset

The CompaCT system is designed to be applied to any medium of DICOM files in the 12-bit grayscale configuration. This study utilized an open-source dataset [22] from The Cancer Imaging Archive (TCIA) of 3954 12-bit DICOM lung CT scans in 512x512 dimensions for evaluating the algorithms. All systems were applied to each CT scan in this dataset, and the resulting compression ratios were calculated.

### 3.4 Evaluation Results

Table 2. Comparisons of industry compression algorithms.

| File | Name | Industry Status | Compression Ratio |
|---|---|---|---|
| Raw | Raw | DICOM Standard | 1 |
| JP2 | JPEG2000-Lossless | DICOM Standard | 1.763210 |
| RLE | Run Length Encoding | DICOM Standard | 1.792206 |
| ZIP | ZIP DEFLATE | DICOM Standard | 1.805547 |
| PNG | Portable Network Graphics | Third-Party | 2.059297 |
| CCT | CompaCT | Proposal | 2.421973 |

To evaluate the enhancements in compression power of the CompaCT format compared to industry standards, we exhibited increased compression ratios against four industry standard and third-party lossless formats: JP2, RLE, ZIP, and PNG as outlined in Table 2. The raw data size is also included as a baseline measurement. JP2, RLE, and ZIP are built-in compressor formats to the DICOM standard package [12]. PNG serves as an external benchmark to evaluate the proposal and is not commonly used in medical image compression.

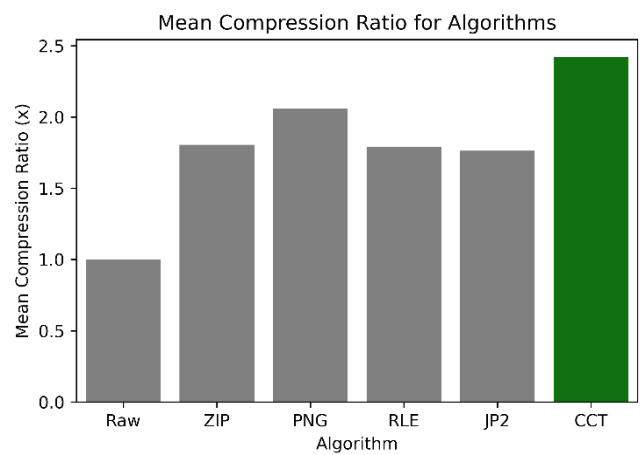

**Fig. 7.** Mean compression ratio for each evaluated algorithm on the TCIA dataset.

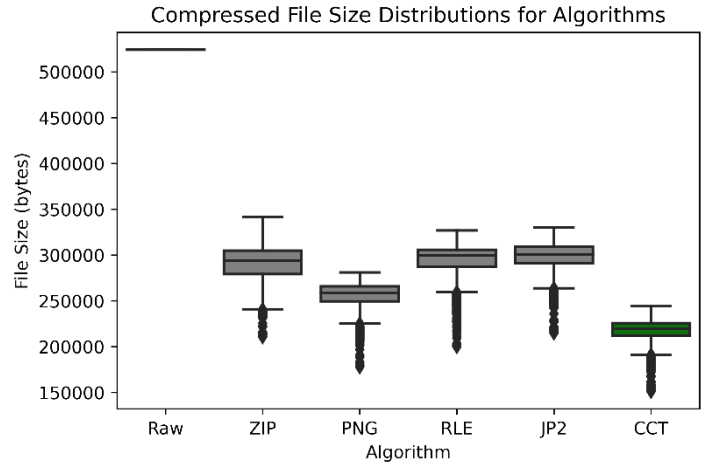

**Fig. 8.** Another view of compression ratio, showing compressed file size distributions for each evaluated algorithm on the dataset.

The CompaCT format results in the largest mean compression ratios compared to all the other evaluated baselines. CompaCT, on average, creates files ~2.42x smaller than the original uncompressed forms as shown in Table 2. To break down the results, CompaCT generates about 37% improved compression ratios compared to the primary DICOM industry standard JP2, and about a 35% improvement compared to RLE as shown in Fig. 7 and 8. CompaCT performs roughly 18% better than PNG and about 34% better than ZIP from Fig. 7 and 8. Compared to the raw uncompressed data, CompaCT is 142% more space-efficient regarding file sizes from Fig. 7 and 8.

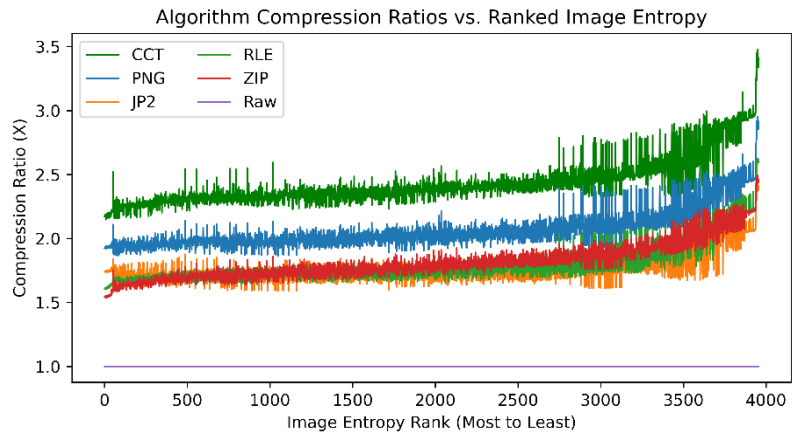

**Fig. 9.** Comparing the compression ratios for each algorithm on each evaluated file, ranked based on entropy present in data (X-axis ranked from hardest to compress to easiest to compress).

Entropy is a measure of randomness present in information. Highly entropic data is more difficult to compress due to fewer repeatable patterns in the information [1].

$$H(x) = -\sum_{i=1}^{n} P(x_i) \log_b P(x_i) \quad (4)$$

Equation (4) defines an estimation of entropy, with $b$ being the base of the logarithm, and $P(x_i)$ being the probability of a certain value, and $n$ being the number of values. The dataset is ranked based on the entropy present in each individual image and plotted for each compressor's power in Fig. 9. CompaCT consistently outperforms the other benchmarks on most images over a broad range of different entropies. In cases with simpler images, CompaCT can optimize compression more drastically than the other systems, proving that the algorithm can generalize and perform on various entropic densities.

### 3.5 Ablation Study

To evaluate the specific components that contributed to the enhancement in compression ratios, a detailed ablation study was conducted. This study involved analyzing the 4 key stages of the pipeline: the Fractal Transform, the Heuristic Pixel Segmentation, the QOI-based Delta Coding, and the DEFLATE compression systems. Each of the components were individually removed from the compression pipeline and the resulting compression ratios were computed on a sample of the CT dataset.

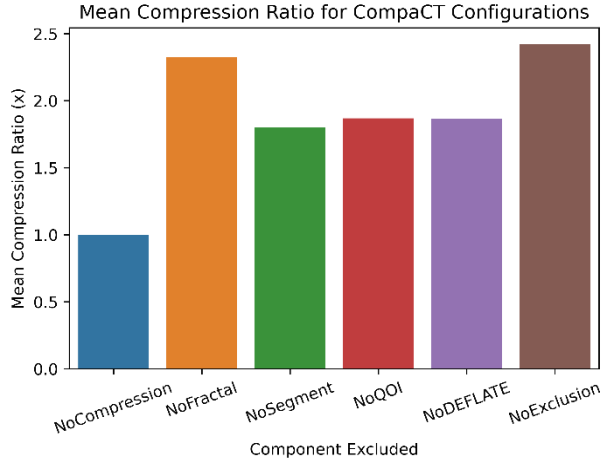

**Fig. 10.** An ablation study of the different components in the CompaCT pipeline. Each labeled component was excluded from the system and the compression ratios were computed. All other steps in the pipeline were kept activated except for the labeled component. "NoCompression" is the baseline without compression, and "NoExclusion" is the overall pipeline together without exclusions.

It is found that in the CompaCT pipeline, removing the segmentation approach results in about a 34% decrease in compression performance. Removing the fractal and reverting to a standard line scan has a 4% decrease in compression. Removing the QOI-based short delta coding results in a 30% decrease in compression, and removing the DEFLATE entropy coding also results in a 30% decrease in compression. It should be noted that the fractal without the segmentation performs relatively worse than the standard raster approach, whereas with the segmentation it outperforms it.

It is intuitive that maximizing the short-delta bit count while staying under a full byte will improve the compression ratios in the QOI-based delta coding. Tightening the window for short deltas to 6, 5, 4, or fewer bits will decrease the count of short deltas that can be cut into a single byte compared to 2 bytes. As shown in the QOI ablation study, removing all short delta coding from the pipeline resulted in significant drops in compression performance. It is in the encoder's best interest to maximize the count of short deltas because each will be equally encoded within a single byte no matter the actual value, it simply needs to meet the short-delta thresholds. For decreasing bitcounts for the short deltas less than 7, there will be a significant drop in performance approaching the results obtained when removing the QOI-short delta coding entirely, implying that 7-bits is optimal when encoding short-deltas and allowing for a single flat bit.

## 4     Discussion

In this study, we introduced a novel lossless image compression codec, CompaCT, designed for optimizing the storage of monochrome high-bit-depth DICOM medical files. Our proposed algorithm demonstrated a 37% increase in compression ratios compared to the current primary DICOM industry standard, JP2. Moreover, the entropic ranking data provided evidence that the compression improvements achieved by CompaCT can be generalized to varying intensities of information. The specific contributions of each component in the CompaCT pipeline were analyzed in a detailed ablation study, concluding that all proposed components in tandem result in the best compression performance for this approach.

The results obtained from our experiments indicate that CompaCT outperforms the current standard algorithms in the DICOM format, resulting in smaller file sizes for compressed images. This validates the efficacy of the enhanced encoder proposed in this study and highlights the potential for utilizing CompaCT to generate smaller dedicated data files of medical images, thereby reducing storage requirements and facilitating more efficient data transfer.

The research presented in this study can hold importance in image processing as it introduces new possibilities for exploring different types of preprocessing and pixel-ordering transforms before traditional encoding systems. The results of CompaCT, a highly specialized segmentation system, paves the way for further research into developing and incorporating other specialized preprocessors to enhance compression algorithms further. This could lead to substantial advancements in image compression technologies and have implications for medical imaging and other domains reliant on efficient data storage and transmission.

While this study focuses primarily on optimizing compression efficiency, it is essential to acknowledge that the gains achieved with the CompaCT algorithm involve a trade-off with compression and decompression speed compared to other existing systems. As the proposal is still under development, we prototyped it in the Python programming language for rapid testing and iteration. However, Python's uncompiled and dynamic nature limits its performance compared to other compiled implementations. Thus, the next logical step in improving this system for future use involves creating an optimized version of the proposed algorithm in a compiled format, ensuring better efficiency and broader applicability. Further optimization can also be achieved by experimenting and utilizing different preprocessing transforms and entropy coding systems, arithmetic coding.

The highly adaptive field of image compression offers numerous opportunities for further research and analysis, contributing to the continuous enhancement of our novel technique of pixel restructuring. The significant influence of pixel image traversal order on compression performance suggests that more optimal orderings within this space may warrant exploration in future studies. Understanding and exploiting these orders could lead to even more substantial gains in compression rates.

Given the results obtained with CompaCT, it is recommended for continued efforts to optimize digital storage through advancements in image compression techniques. Further studies can explore how this algorithm performs with different types of medical images, datasets, and applications to validate its robustness and versatility. Additionally, investigating its adaptability to color images and extending its capabilities to lossy compression scenarios could open new avenues for broader adoption across various industries.

## 5   Conclusion

Our findings demonstrate the efficacy of the CompaCT algorithm in enhancing compression ratios for medical images. Through strategic pixel restructuring, coupled with innovative delta coding and entropy coding techniques, CompaCT achieved a 37% increase in compression ratios compared to the current industry standard, JP2. This achievement underscores the algorithm's ability to transform the storage landscape for medical images, minimizing costs and facilitating seamless data sharing.

The broader significance of this study extends beyond medical image compression. The results of CompaCT showcase the potential of specialized algorithms to outperform general-purpose compression methods. The approach of utilizing pixel restructuring, delta coding, and entropy coding can inspire advancements in compression techniques applicable across diverse domains where data storage and transmission efficiency are paramount.

As medical imaging continues to evolve and play an increasingly central role in healthcare, the optimized storage and efficient transmission of medical images become imperative. By addressing these avenues for further research, this study can drive the development of efficient and powerful compression algorithms that will play a pivotal role in shaping the future of data storage and transmission.

# Abbreviations

| Abbreviation | Definition |
|---|---|
| CCT | CompaCT |
| CLD | Count of Large Differences |
| CT | Computed Tomography |
| DICOM | Digital Imaging and Communications in Medicine |
| JP2 | JPEG2000-Lossless |
| L | Large Difference |
| LZ77 | Lempel-Ziv 1977 |
| MRI | Magnetic Resonance Imaging |
| PNG | Portable Network Graphics |
| QOI | Quite OK Image Format |
| RLE | Run Length Encoding |
| RMSE | Root Mean Square Error |
| ROI | Region of Interest |
| TCIA | The Cancer Imaging Archive |

# Declarations

**Availability of Data and Materials**

The dataset used to evaluate the current study are available in The Cancer Imaging Archive repository, https://doi.org/10.7937/K9/TCIA.2015.NPGZYZBZ [22]. The code used and analyzed during the current study is available as an open-source software package in the repository at https://doi.org/10.5281/zenodo.8274859. The latest version of the CompaCT project is available at https://github.com/taaha-khan/2023-CompaCT-Image-Compression under the CC BY-NC-SA 4.0 license. The operating system is platform independent, with requirements of Python 3.8 or higher.

**Competing Interests**

The authors declare that they have no competing interests.

**Funding**

Not applicable

**Authors' Contributions**

The idea, implementation, and validation of this work was conducted by the main author TK. The author wrote, read, and approved the final manuscript.

**Acknowledgements**

Special appreciation to F. M. Kashif and B. Zia for providing feedback and comments on the manuscript.

15. J Moreno, X Otazu, Image compression algorithm based on Hilbert Scanning of Embedded quadTrees: an introduction of the Hi-SET coder. In Proceedings of the 2011 IEEE International Conference on Multimedia and Expo, 11-15 July 2011
16. S Biswas, One-dimensional B-B polynomial and Hilbert scan for graylevel image coding. Pattern Recognition 37, 789-800 (2004). doi:10.1016/j.patcog.2003.09.001.
17. S Amraee, N Karimi, S Samavi, S Shirani, Compression of 3D MRI images based on symmetry in prediction-error field. In Proceedings of the 2011 IEEE International Conference on Multimedia and Expo, 1-15 July 2011
18. R Kumar MS, S Koliwad, G Dwarakish, Lossless compression of digital mammography using fixed block segmentation and pixel grouping. In Proceedings of the 2008 Sixth Indian Conference on Computer Vision, Graphics & Image Processing, 16-19 December 2008
19. J Cerveny, Gilbert generalized Hilbert source code Version 1.0. (Github, 2021), https://github.com/jakubcerveny/gilbert. Accessed 30 Dec 2022.
20. D Saupe, Optimal piecewise linear image coding, Visual Communications and Image Processing 3309, 747-760 (1998). doi:10.1117/12.298387.
21. D Szablewski, Lossless image compression in O(N) time. (Phoboslab, 2021), https://phoboslab.org/log/2021/11/qoi-fast-lossless-image-compression. Accessed 30 Dec 2022.
22. D Goldgof et al., Qin lung CT Version 2 (The Cancer Imaging Archive, 2015), doi:10.7937/K9/TCIA.2015.NPGZYZBZ.
23. BH Menze et al., The multimodal brain tumor image segmentation benchmark (brats), IEEE Transactions on Medical maging 34(10), 1993-2024 (2015). doi: 10.1109/tmi.2014.2377694
24. S Bakas et al., Identifying the best machine learning algorithms for brain tumor segmentation, progression assessment, and overall survival prediction in the BRATS challenge. ArXiv (2018). doi:10.48550/arXiv.1811.02629
25. S Bakas et al., Advancing the cancer genome Atlas Glioma MRI collections with expert segmentation labels and Radiomic features. Nature Scientific Data 4(1), (2017). doi: 10.1038/sdata.2017.117.